\documentclass[aps,prb,twocolumn,superscriptaddress,floatfix]{revtex4-1}
\usepackage{bm}
\usepackage{amsmath}
\usepackage{amssymb}
\usepackage{graphicx}
\usepackage{color}
\usepackage{soul,xcolor}
\setstcolor{red}
\usepackage{hyperref}

\newcommand{\kB}{k_{\text{B}}}
\newcommand{\EF}{E_{\text{F}}}
\newcommand{\deriv}[2]{\frac{\text{d}{#1}}{\text{d}{#2}}}

\begin{document}

\title{Strain-induced enhancement of the Seebeck effect in magnetic tunneling junctions via interface resonant tunneling: \textit{Ab-initio} study}

\author{Kaoru Yamamoto}
\affiliation{Research Center for Magnetic and Spintronic Materials, National Institute for Materials Science (NIMS), Tsukuba 305-0047, Japan}
\email{YAMAMOTO.Kaoru@nims.go.jp}

\author{Keisuke Masuda}
\affiliation{Research Center for Magnetic and Spintronic Materials, National Institute for Materials Science (NIMS), Tsukuba 305-0047, Japan}

\author{Ken-ichi Uchida}
\affiliation{Research Center for Magnetic and Spintronic Materials, National Institute for Materials Science (NIMS), Tsukuba 305-0047, Japan}
\affiliation{Research and Services Division of Materials Data and Integrated System, National Institute for Materials Science (NIMS),Tsukuba 305-0047, Japan}
\affiliation{Institute for Materials Research, Tohoku University, Sendai 980-8577, Japan}
\affiliation{Center for Spintronics Research Network, Tohoku University, Sendai 980-8577, Japan}
\affiliation{Department of Mechanical Engineering, The University of Tokyo, Tokyo 113-8656, Japan} 

\author{Yoshio Miura}
\affiliation{Research Center for Magnetic and Spintronic Materials, National Institute for Materials Science (NIMS), Tsukuba 305-0047, Japan}
\affiliation{Research and Services Division of Materials Data and Integrated System, National Institute for Materials Science (NIMS), Tsukuba 305-0047, Japan}
\affiliation{Center for Spintronics Research Network, Osaka University, Toyonaka, Osaka 560-8531, Japan}

\date{\today}% It is always \today, today,

\begin{abstract}
We investigate the thermoelectric properties of Fe/MgO/Fe(001) magnetic tunnel junctions (MTJs) by means of the linear-response theory combined with a first-principles-based Landauer-B\"uttiker approach. 
We find that the Seebeck coefficient of Fe/MgO/Fe(001) MTJs strongly depends on the barrier thickness and the tetragonal distortion. 
A compressive tetragonal distortion of the in-plane lattice parameter in the MTJs provides interface resonant states just above the Fermi energy. 
This causes resonant tunneling in the MTJs and significantly enhances the Seebeck coefficient when the thickness of the MgO barrier is around 1 nm (four or five atomic layers of MgO).
Moreover, an extensive tetragonal distortion of the in-plane lattice parameter pushes the interface states away from the Fermi energy, leading to a reduction of the Seebeck coefficient. 
Furthermore, we find that the interface resonant tunneling enhances the power factor of the MTJs for the compressive distortion. 
These results indicate that control of the barrier thickness and the tetragonal distortion will be effective for maximizing the thermoelectric properties of MTJs.
\end{abstract}

\maketitle

\section{Introduction}

Since initial observations of the spin Seebeck effect \cite{Uchida2008Nature, Uchida2010NatMaterspinSeebeck, Jaworski2010NatMaterspinSeebeck, Uchida2010APLspinSeebeck}, the conversion between spin and heat currents has been actively investigated, which has opened up an emerging research area called spin caloritronics \cite{Bauer2012NatMaterReview}. 
This area encompasses novel thermoelectric phenomena in various magnetic systems ranging from bulk magnets to magnetic tunneling junctions (MTJs). 

The heat-to-charge current conversion, namely, the Seebeck effect, in MTJs has been recently investigated both experimentally \cite{Liebing2011PRLCofeBTpower, Walter2011NatMaterFeCo, Lin2012NatCommnAl2O3, Zhang2012PRLCoFeB, Boehnke2015SciRepbias, Shan2015PRBCoFeBMgO, Heubner2016PRBCOFeBMgAl2O4, Boehnke2017NatCommnHeusler, Heubner2017PRBCoFeBMgAl2O4, Ning2017AIPAdvCoFeBMgO, Heubner2018JphysDCoFeBMAO} and theoretically \cite{Czerner2011PRBFeCo, Czerner2012JAPCoFe, Heliger2013PRBFeCo, Amin2014PRBCoPt, Lopez-Monis2014PRBSonczewskiModel, Comtesse2014PRB, Wang2014PRBFeCo, Geisler2015PRBHeusler, Jia2016NJPFeMgO, Li2019PCCPHeuslerBoltz}, in which most studies have considered the tunnel magneto-Seebeck (TMS) ratio \cite{Kuschel2019JphysDReview}, the thermoelectric analog of the tunnel magnetoresistance ratio.
MTJs with a large Seebeck effect have potential not only to produce a large TMS ratio, but also to realize various applications such as thermal energy harvesting, thermoelectric cooling in nanoscale spintronic devices \cite{Kuschel2019JphysDReview}, and scanning Seebeck tunneling microscopy \cite{Friesen2018JPysDscan, Friesen2019Science}.

According to the linear-response theory, a large Seebeck effect can be obtained through an asymmetric energy dependence of the electronic transport with respect to the Fermi energy \cite{Shakouri2011AnnuRevMaterReview}.
In MTJs, electronic structures not only in the bulk electrode but also in the interface can provide an asymmetric energy dependence of the tunneling conductance.
For example, recent theoretical studies \cite{Comtesse2014PRB, Geisler2015PRBHeusler} have tried to understand the Seebeck effect in MTJs from the density of states (DOS) of ferromagnetic electrodes.
In particular, Boehnke \textit{et al.}~\cite{Boehnke2017NatCommnHeusler} have investigated various ferromagnetic materials for electrodes in MTJs that have a preferable DOS for a large Seebeck effect.
They have experimentally observed a relatively large Seebeck effect in MTJs such as $\text{Co}_2\text{FeAl}\text{/MgO/CoFeB}$ and $\text{Co}_2\text{FeSi}\text{/MgO/CoFe}$.

While these investigations have focused only on the bulk electronic structure of the ferromagnetic electrode, the interface property is also an important factor for the Seebeck effect in MTJs.
For example, the interfacial termination has been found to affect the Seebeck effect in MTJs \cite{Czerner2012JAPCoFe, Wang2014PRBFeCo, Geisler2015PRBHeusler}.
Furthermore, Jia \textit{et al.}~\cite{Jia2016NJPFeMgO} have reported that the interface resonant tunneling significantly changes the energy dependence of the tunneling conductance.
They have calculated the Seebeck coefficient of Fe/MgO/Fe(001) MTJs, which are known to exhibit interface resonant tunneling due to the large DOS of minority-spin states at the interface \cite{Bulter2001PRB, Mathon2001PRB, Belashchenko2005PRBIRS}.
While the enhancement of the Seebeck coefficient due to the interface resonant tunneling has been pointed out, detailed properties of the interface resonant tunneling have not been discussed in their work \cite{Jia2016NJPFeMgO}. 
In particular, it is expected that the interface resonant tunneling strongly depends on the structural parameters of MTJs, such as the barrier thickness and lattice distortion.

In the present work, we theoretically investigate the Seebeck effect in the Fe/MgO/Fe(001) MTJ by using the linear-response theory combined with a first-principles-based Landauer-B\"uttiker approach while changing the barrier thickness and the tetragonal distortion of the in-plane lattice parameter. 
Since the lattice mismatch between bcc Fe and MgO is 4\%, the Fe/MgO/Fe(001) MTJ has a tetragonal distortion depending on the in-plane lattice constant.
Such a distortion is expected to change the interfacial electronic structures and to affect the interface resonant tunneling.
Here, we use the following two kinds of in-plane lattice constants: one is the experimental value of the lattice constant of bcc Fe ($a=a_\text{Fe}=2.866$~\AA) and the other is that of MgO ($a=a_\text{MgO}/\sqrt{2}=2.987$~\AA).
These compressive and extensive tetragonal distortions of the in-plane lattice parameter can be potentially realized in experiments by changing the material of the buffer layer or thickness of the ferromagnetic electrode.
For example, the in-plane lattice constant of the Fe/MgO/Fe(001) MTJ might approach that of bcc Fe by increasing the thickness of the Fe electrode.
We found that these in-plane lattice distortions significantly influence the energy difference between bonding and antibonding states of interfacial Fe atoms around the Fermi level and affect the Seebeck coefficient of the MTJs.
Furthermore, we found that the interface resonant tunneling also enhances the power factor of the MTJ for the compressive distortion.

\section{Calculation procedure}

\begin{figure}
    \centering
    \includegraphics[width=0.9\linewidth]{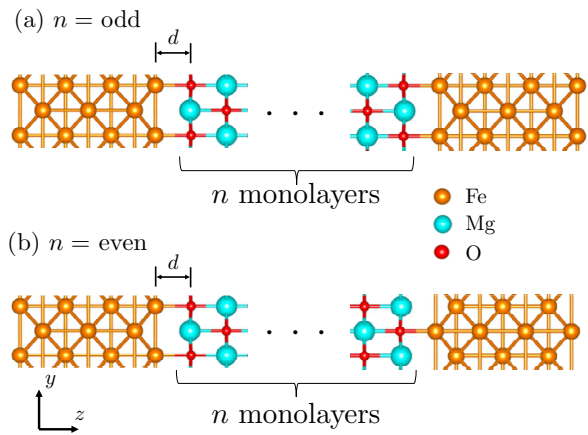}
    \caption{Supercells of the Fe(7ML)/MgO($n$ML)/Fe(7ML)(001) MTJ with (a) an odd and (b) an even number of MgO layers. 
    The atomic configuration at the right interface is different between the two cases.}
    \label{fig:structure}
\end{figure}

\subsection{Model}
We prepared supercells of Fe(7ML)/MgO($n$ML)/Fe(7ML)(001) MTJs for several numbers of MgO layers $n$ (from $n=3$ to $n=12$) (Fig.~\ref{fig:structure}) and for two kinds of tetragonal distortions, namely, the compressive tetragonal distortion of the in-plane lattice parameter ($a=a_\text{Fe}=2.866 \text{ \AA}$) and the extensive tetragonal distortion of the in-plane lattice parameter ($a=a_\text{MgO}/\sqrt{2}=2.978\text{ \AA}$).
Since stable interface structures are required for correct estimation of the Seebeck coefficient, we carried out the structural optimization for each supercell by using the density-functional theory (DFT) combined with the generalized gradient approximation for exchange-correlation energy \cite{Perdew1996PRLGGA}, which was implemented in the Vienna \textit{Ab-initio} Simulation Package (VASP) \cite{Kresse1996PRBvasp1, Kresse1999PRBvasp2}.
The details of the structural optimization are given in our previous work \cite{Masuda2017PRB}.
A $15 \times 15 \times 1$ \textbf{k}-point mesh was used in the Brillouin-zone integrations for all the supercells.
As a result of the calculations, we found that the interface distances were approximately $2.2 \text{ \AA}$ for the compressive distortion ($a=a_\text{Fe}=2.866~\text{\AA}$), and $2.1 \text{ \AA}$ for the extensive distortion ($a=a_\text{MgO}/\sqrt{2}=2.978$~\AA). 
The difference in the interface distance between the two cases significantly affects the interface resonant tunneling and the Seebeck effect in MTJs.

\subsection{Calculation of thermoelectric parameters}
The electric current in MTJs is given by the Landauer-B\"uttiker formula \cite{Datta1997} as
\begin{equation}
I = \frac{e}{h}\int  dE  \ \tau(E) [f_\text{L}(E)-f_\text{R}(E)], \label{eq:I}  
\end{equation}
where $e  (>0)$ is the elemental charge, $h$ is the Plank constant, $\tau(E) = \tau_\uparrow(E)+\tau_\downarrow(E)$ is the total transmittance in the MTJ with $\tau_\uparrow(E)$ [$\tau_\downarrow(E)$] being the transmittance in the majority-spin (minority-spin) channel, and $f_\text{L(R)} = [1+e^{\beta_\text{L(R)}(E-\mu_\text{L(R)})}]^{-1}$ is the Fermi function with the inverse temperature $\beta_\text{L(R)}={(\kB T_\text{L(R)})}^{-1}$ and the chemical potential $\mu_\text{L(R)}$ of the left (right) Fe electrode.
Here, $\kB$ is the Boltzmann constant and $T_\text{L(R)}$ is the temperature of the left (right) Fe electrode.
We calculated the transmittance by means of the DFT \cite{BrandbygePRB2002ATK1}, which was implemented in the Atomistix ToolKit package (ATK) \cite{ATK, Smidstrup2017PRBATK2,Smidstrup2019JphysCondATK3}; see Refs.~\onlinecite{BrandbygePRB2002ATK1,Masuda2017PRB} for further details on the calculations.
In the DFT calculations implemented in the ATK, we utilized the double-$\zeta$ polarized basis set.
A $7 \times 7 \times 50$ \textbf{k}-point mesh was used for all of the cases except for the compressive distortion ($a=a_\text{Fe}=2.866~\text{\AA}$) with $n=12$ ($7\times7\times70$ \textbf{k}-point mesh) and for the extensive distortion ($a=a_\text{MgO}/\sqrt{2}=2.978~\text{\AA}$) with $n=10$ ($10\times10\times150$ \textbf{k}-point mesh) because of the poor convergence for these two cases with the $7 \times 7 \times 50$ \textbf{k}-point mesh.
Such a large number of $k_z$ points are required to minimize the mismatch of the Fermi energy between the Fe electrode and scattering region including the barrier, where an open boundary condition is adopted for the transport calculation implemented in the ATK \cite{Smidstrup2019JphysCondATK3, kzpoint}.
We confirmed good numerical convergence with the appropriate \textbf{k}-point meshes above for all of the cases.

In the transport calculations, we set the in-plain \textbf{k} points as $\textbf{k}_{\parallel}=801\times 801$ for all of the cases.
These many in-plane \textbf{k} points are needed to evaluate some sharp peaks in the $\textbf{k}_\parallel$ dependence of the transmittance, which appear in MTJs with the interface resonant tunneling \cite{Bulter2001PRB, Mathon2001PRB}.

The linear expansion of the electric current with respect to the temperature and the chemical-potential differences ($\Delta T = T_\text{L}-T_\text{R}$ and $\Delta \mu = \mu_\text{L}-\mu_\text{R}$) gives the electric conductance as
\begin{equation}
G \equiv \left(\deriv{I}{V}\right)_{\Delta T=0} = e^2L_0, \label{eq:G}
\end{equation}
and the Seebeck coefficient as 
\begin{equation}
S \equiv -\left(\frac{\Delta \mu/e}{\Delta T}\right)_{I=0} =  -\frac{1}{eT}\frac{L_1}{L_0}, \label{eq:Seebeck}
\end{equation}
where we set $T_\text{L}=T_\text{R}=T$ in the linear-response regime.
Here, $L_p \ (p=0,1)$ is the generalized transport coefficient defined as
\begin{equation}
L_p \equiv \frac{1}{h}\int_{-\infty}^{\infty}dE \  (E-\EF)^p \tau(E)\left(-\deriv{f}{E}\right), \label{eq:L_n}
\end{equation}
where $\EF$ is the Fermi energy and $f = [1+e^{\beta(E-E_\text{F})}]^{-1}$, with $\mu_\text{L}=\mu_\text{R}=E_\text{F}$ and $\beta_\text{L}=\beta_\text{R}=\beta$.
Here, we neglect the temperature dependence of the chemical potential and assume $\mu = E_\text{F}$.
Throughout the present work, the temperature $T$ in the Fermi distribution function is fixed to $300 \text{ K}$.

\section{Results and discussion}
\begin{figure}
    \centering
    \includegraphics[width=0.9\linewidth]{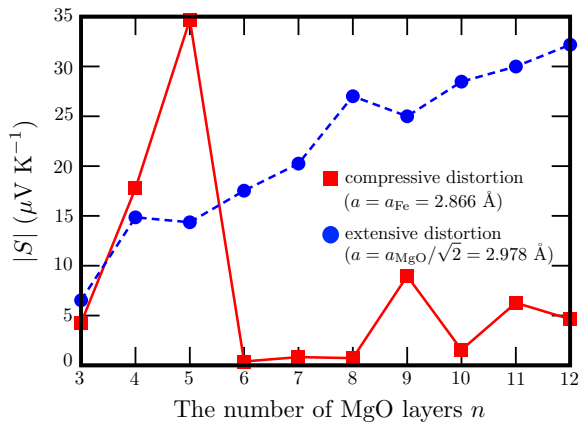}
    \caption{The absolute value of the Seebeck coefficient of the Fe(7ML)/MgO($n$ML)/Fe(7ML)(001) MTJ for the parallel magnetization configuration as a function of the number of MgO layers $n$ for the compressive ($a=a_\text{Fe}=2.866\text{ \AA}$, red solid curve) and the extensive ($a=a_\text{MgO}/\sqrt{2}=2.978\text{ \AA}$, blue dashed curve) distortions.}
    \label{fig:Seebeckpara}
\end{figure}

Figure~\ref{fig:Seebeckpara} shows the absolute value of the Seebeck coefficient of the Fe(7ML)/MgO($n$ML)/Fe(7ML)(001) MTJ for the parallel magnetization configuration as a function of the number of MgO layers $n$ for the compressive ($a=a_\text{Fe}=2.866 \text{ \AA}$) and the extensive ($a=a_\text{MgO}/\sqrt{2}=2.978 \text{ \AA}$) tetragonal distortions of the in-plane lattice parameter.
In the present work, we mainly analyzed the Seebeck coefficient for the parallel magnetization configuration of the MTJ, in which we can clearly see the effect of the interface resonant tunneling on the Seebeck effect.
We give the analysis of the antiparallel magnetization configuration in the Appendix, in which we can see the effect that might be due to the interface resonant tunneling. 
The behavior of the Seebeck coefficient is completely different between the two cases.
While the Seebeck coefficient for the extensive distortion increases almost gradually and monotonically as $n$ increases, the one for the compressive distortion reaches its maximum at $n=5$.

The signs of the Seebeck coefficients are all negative except for $n=6$ and $7$ for the compressive distortion.
The absolute values of the Seebeck coefficient are approximately tens of $\mu \text{V} \text{K}^{-1}$, and these are consistent with previous theoretical work \cite{Czerner2011PRBFeCo, Heliger2013PRBFeCo, Jia2016NJPFeMgO} and experimental work with CoFeB electrodes \cite{Heubner2017PRBCoFeBMgAl2O4}, but the results are about one to two orders smaller than those observed in the other previous experiments with CoFeB electrodes \cite{Walter2011NatMaterFeCo, Liebing2012JAPCoFeBMgO, Boehnke2013RevSciInstCoFeB, Boehnke2015SciRepbias, Heubner2016PRBCOFeBMgAl2O4, Boehnke2017NatCommnHeusler, Ning2017AIPAdvCoFeBMgO}.
This discrepancy might originate from the estimation of the temperature difference in the experiments as pointed out in Refs.~\onlinecite{Zhang2015PRLphonon, Jia2016NJPFeMgO}.

We now focus on the following four characteristic behaviors of the Seebeck coefficient found in Fig.~\ref{fig:Seebeckpara}: (1) the enhancement of the Seebeck coefficient for the compressive distortion for $n=4$ and $5$ and the relatively small Seebeck coefficients for the other $n$, (2) the difference in the Seebeck coefficient between the compressive and the extensive distortions for $n=4$ and $5$, (3) the oscillation of the Seebeck coefficient on the odd and even $n$ for the compressive distortion, and (4) the gradual increase in the Seebeck coefficient for the extensive distortion as $n$ increases.

\begin{figure}
    \centering
    \includegraphics[width=0.9\linewidth]{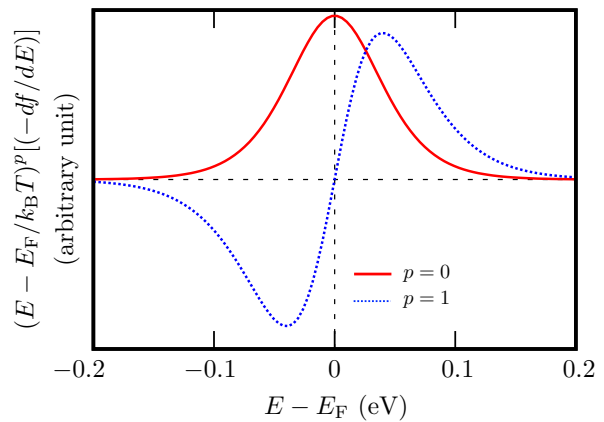}
    \caption{Plots of $-(\text{d}f/\text{d}E)$ and $(E-\EF)[ -(\text{d}f/\text{d}E)]$ normalized by $\kB T$ at room temperature, $T=300\text{ K}$, as a function of energy relative to the Fermi energy $E-\EF$, which are included in the integrand of the generalized transport coefficients [Eq.~\eqref{eq:L_n}].}
    \label{fig:FermiDeriv}
\end{figure}

In order to investigate these behaviors, we analyzed the energy dependence of the transmittance since it dominates the behavior of the Seebeck coefficient via the generalized transport coefficients in Eq.~\eqref{eq:L_n}. 
The Seebeck coefficient increases with decreasing $L_0$ and increasing $L_1$ as shown in Eq.~\eqref{eq:Seebeck}. 
A small $L_0$, however, results in a small conductance, thus leading to low electric power.
For example, an insulator usually has a large Seebeck coefficient but it is not useful for thermoelectric applications.
We therefore focus on how to enhance $L_1$ for a large Seebeck coefficient.
Since $(E-\EF)[-(\text{d}f/\text{d}E)]$ appearing in the integrand in $L_1$ is an antisymmetric function with respect to the Fermi energy, $L_1$ vanishes with a symmetric energy dependence of transmittance with respect to the Fermi energy, thus resulting in a zero Seebeck coefficient.
Therefore, we need an asymmetric energy dependence of transmittance with respect to the Fermi energy for a larger $L_1$.
Moreover, such an asymmetric energy dependence should appear near the Fermi energy since $(E-\EF)[-(\text{d}f/\text{d}E)]$ is almost zero at the energy far away from the Fermi energy (Fig.~\ref{fig:FermiDeriv}).
At 300 K, the asymmetric energy dependence of the total transmittance appearing in the energy range $E-\EF = [-0.2 \text{ eV}, 0.2 \text{ eV}]$ contributes to the Seebeck coefficient.

\begin{figure}
    \centering
    \includegraphics[width=\linewidth]{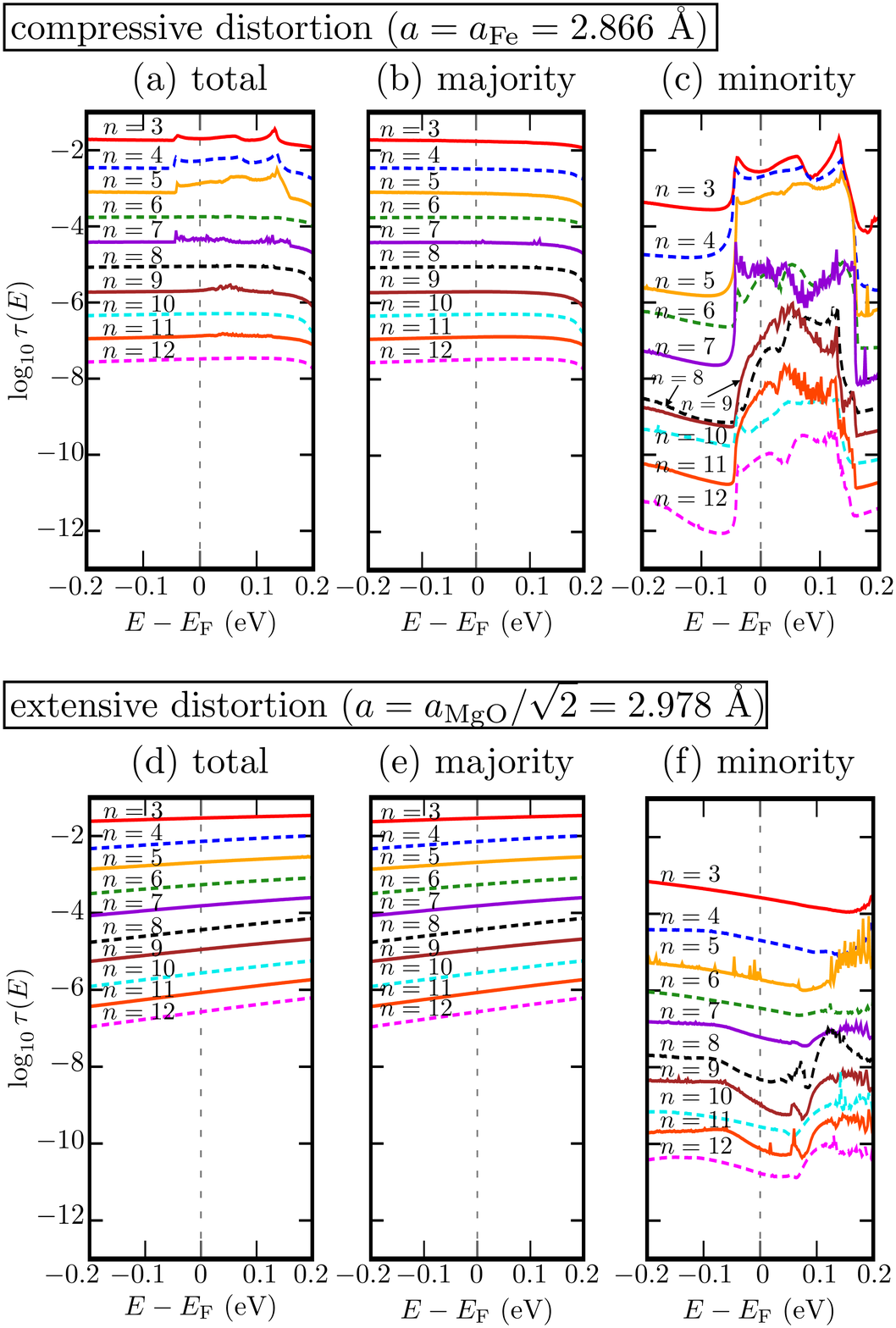}
    \caption{(a) Total transmittance, (b) transmittance in the majority-spin channel, and (c) that in the minority-spin channel in the Fe(7ML)/MgO($n$ML)/Fe(7ML)(001) MTJ for the parallel magnetization configuration as a function of energy relative to the Fermi energy $E-\EF$ for the number of MgO layers $n$ from $n=3$ to $n=12$ and the compressive distortion ($a=a_\text{Fe}=2.866 \text{ \AA}$). 
    (d)-(f) The same as (a)-(c), but for the extensive distortion ($a=a_\text{MgO}/\sqrt{2}=2.978 \text{ \AA}$). 
    The solid curves show the transmittances for odd $n$, while the dashed curves show those for even $n$. 
    We plotted the data in the energy range $E-\EF = [-0.2 \text{ eV}, 0.2 \text{ eV}]$, which mainly contributes to the Seebeck coefficients at $T=300$ K since the factor $(E-\EF)[-(\text{d}f/\text{d}E)]$ at $T=300$ K in Fig.~\ref{fig:FermiDeriv} is almost zero outside the range.}
    \label{fig:Transpara}
\end{figure}

\begin{figure}
    \centering
    \includegraphics[width=\linewidth]{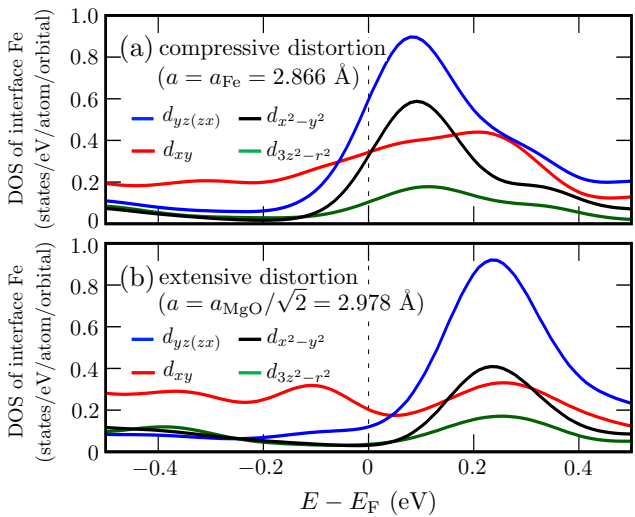}
    \caption{The DOS of the minority-spin $d$ states of the interface Fe atoms as a function of energy relative to the Fermi energy $E-\EF$ for (a) the compressive ($a=a_\text{Fe}=2.866\text{ \AA}$) and (b) the extensive ($a=a_\text{MgO}/\sqrt{2}=2.978$ \AA) distortions.}
    \label{fig:LDOS}
\end{figure}

We first consider the enhancement of the Seebeck coefficient for the compressive distortion for $n=4$ and $5$ and the relatively small Seebeck coefficients for the other $n$ in Fig.~\ref{fig:Seebeckpara}.
Figure \ref{fig:Transpara}(a) shows the total transmittance for the compressive distortion as a function of energy relative to the Fermi energy.
For $n=4$ and $5$, the strong enhancement appears in the total transmittance in the energy range $E-\EF = [-0.05 \text{ eV}, 0.14 \text{ eV}]$.
This makes the energy dependence of the transmittance asymmetric, thus leading to the enhancement in the Seebeck coefficient for the compressive distortion for $n=4$ and $5$ via the increase in $L_1$.
The enhancement in the total transmittance comes from the transmittance in the minority-spin channel shown in Fig.~\ref{fig:Transpara}(c), which is due to the interface resonant tunneling between the left and right interfaces.
Figure \ref{fig:LDOS}(a) shows the DOSs in the minority-spin $d$ states at the interface Fe atoms as a function of energy relative to the Fermi energy for the compressive distortion. 
We find the peak of DOS of $d_{yz(zx)}$ and $d_{x^2-y^2}$ states centered around $E-E_\text{F} = 0.1~\text{eV}$, which mainly contributes to the interface resonant tunneling.

For the other numbers of MgO layers ($n = 3$, $n \geq 6$), the enhancement in the total transmittance due to the interface resonant tunneling is not as clear as that for $n=4$ and $5$, thus resulting in small Seebeck coefficients. 
For $n\geq 6$, we can not see a clear enhancement in the total transmittance as shown in Fig.~\ref{fig:Transpara}(a) because the wave function of the interface resonant states in the minority-spin state shows the fast decay in the MgO barrier as compared with that of the majority-spin $\Delta_1$ evanescent state. 
For $n=3$, the enhancement due to the interface resonant tunneling in the total transmittance appears but it is smaller around the Fermi energy than that for $n=4$ and $5$, leading to a smaller value of the Seebeck coefficient than these cases. 
The reason why the interface resonant tunneling is relatively suppressed for $n=3$ is that the thickness of the barrier is too small to describe the insulating behavior of MgO.
Other states such as $\Delta_5$ and $\Delta_2$ states contribute to the transmittance in the majority-spin channel and smear the enhancement of the transmittance due to the interface resonant tunneling.
Therefore, the transmittance in the majority-spin channel shown in Fig.~\ref{fig:Transpara}(b) dominates the total transmittance for $n=3$ and $n\geq 6$, thus leading to the almost symmetric energy dependence of the total transmittance with respect to the Fermi energy. 
The generalized transport coefficient $L_1$ thus becomes almost zero, resulting in an almost zero Seebeck coefficient.

\begin{figure}
    \centering
    \includegraphics[width=\linewidth]{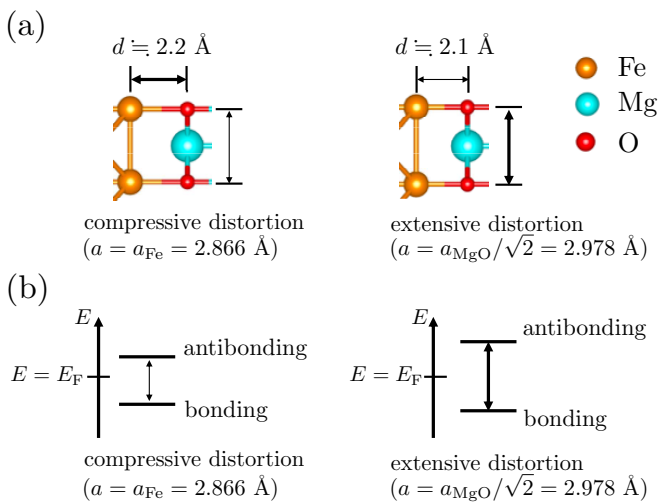}
    \caption{(a) A schematic picture of the left interface of the Fe/MgO/Fe(001) MTJ for the compressive ($a=a_\text{Fe}=2.866\text{ \AA}$) and the extensive ($a=a_\text{MgO}/\sqrt{2}=2.978$~\AA) distortions. 
    As the in-plane lattice constant increases, the interface distance decreases. 
    (b) A schematic picture of the bonding and antibonding states at the interface for the compressive ($a=a_\text{Fe}=2.866\text{ \AA}$) and the extensive ($a=a_\text{MgO}/\sqrt{2}=2.978$~\AA) distortions. 
    As the interface distance decreases, the coupling at the interface becomes strong, thus leading to an increase in the energy difference between the bonding and the antibonding states.}
    \label{fig:interface}
\end{figure}

Next, we consider the difference in the Seebeck coefficient between the compressive and the extensive distortions for $n=4$ and $5$ found in Fig.~\ref{fig:Seebeckpara}.
In contrast to the case for the compressive distortion shown in Fig.~\ref{fig:Transpara}(a), the enhancement in the total transmittance due to the interface resonant tunneling does not appear for the extensive distortion as shown in Fig.~\ref{fig:Transpara}(d).
This is because the maximal transmittance caused by the interface resonant tunneling shifts to the higher-energy side while increasing the in-plane lattice constant [see the transmittances in the minority-spin channel for both cases shown in Figs.~\ref{fig:Transpara}(c) and \ref{fig:Transpara}(f)].
In this case, the maximal transmittance from the interface resonant tunneling does not contribute to the Seebeck coefficient because the factor $(E-\EF)[-(\text{d}f/\text{d}E)]$ at $T=300$ K does not pick up the maximal transmittance from the interface resonant tunneling outside the energy range $E-\EF = [-0.2 \text{ eV}, 0.2 \text{ eV}]$.
Therefore, the Seebeck coefficient for the extensive distortion is smaller than that for the compressive distortion. 
In order to understand the shift of the maximal transmittance caused by the interface resonant tunneling, we calculated the local DOS of interfacial Fe atoms at the Fe/MgO(001) interface for the extensive distortion of the in-plane lattice shown in Figs.~\ref{fig:LDOS}(a) and \ref{fig:LDOS}(b).
We confirm that the interfacial states of Fe $d_{yz(zx)}$ orbitals are shifted to the higher-energy side due to the extensive distortion and are located around $E-\EF > 0.2 \text{ eV}$, which is consistent with the shift of the maximal transmittance from the interface resonant tunneling.
These interfacial states are mainly composed of antibonding states of interfacial Fe $d_{yz(zx)}$ orbitals.
Since the interfacial Fe-O distance decreases due to the extensive distortion of the in-plane lattice parameter as shown in Fig.~\ref{fig:interface}(a), the energy difference between the bonding and antibonding states increases with the distortion [see Fig.~\ref{fig:interface}(b)].
This gives the shift of the interfacial antibonding states to the higher-energy side, thus leading to a smaller contribution to the Seebeck coefficient.

The oscillation of the Seebeck coefficient on the odd and even $n$ for $n \geq 5$ for the compressive distortion found in Fig.~\ref{fig:Seebeckpara} is also due to the interface resonant tunneling.
For odd $n$, the enhancement of the total transmittance due to the interface resonant tunneling shown in Fig.~\ref{fig:Transpara}(a) gives larger Seebeck coefficients than those for even $n$. 
For even $n$, the enhancement of the total transmittance does not appear as shown in Fig.~\ref{fig:Transpara}(a), thus leading to smaller Seebeck coefficients with almost symmetric transmittance from the majority-spin $\Delta_1$ evanescent state.
The different behavior of the enhancement in the total transmittance due to the interface resonant tunneling between even and odd $n$ causes the oscillation of the Seebeck coefficient.
This may originate from the symmetry of the left and right interfacial structures, which are symmetric for odd $n$ but asymmetric for even $n$ since the atomic positions of Mg and O atoms are shifted as shown in Fig.~\ref{fig:structure}.
For symmetric interfacial structures with odd $n$, the effect of the interface resonant tunneling is noticeable as compared with that for asymmetric interfacial structures with even $n$, thus resulting in the enhancement in the total transmittance.
For $n=7$, however, the Seebeck coefficient is almost zero because the enhancement due to the interface resonant tunneling appears almost symmetrically in the total transmittance.

A gradual increase in the Seebeck coefficient with increasing $n$ for the extensive distortion was found in Fig.~\ref{fig:Seebeckpara}.
This comes from the increase in the gradient in the total transmittance as $n$ increases [Fig.~\ref{fig:Transpara}(d)].
The increase in the gradient makes the total transmittance more asymmetric and enhances the Seebeck coefficient via the increase in $L_1$.
We found the following properties of the increase in the gradient in the energy dependence of the transmittances.
First, the increase in the gradient for the extensive distortion with increasing $n$ comes from that in the transmittance in the majority-spin channel coming from the $\Delta_1$ state [Fig.~\ref{fig:Transpara}(e)] since it dominates the transport for the extensive distortion. 
Second, for the same number of MgO layers, the gradient increases with the in-plane lattice constant. 
This can be found by comparing the transmittances in the majority-spin channel for the compressive and the extensive distortions at fixed $n$, as shown in Figs.~\ref{fig:Transpara}(b) and \ref{fig:Transpara}(e).
Third, for the fixed in-plane lattice constant, the gradient increases with increasing $n$ even for the compressive distortion when the transmittance in the majority-spin channel dominates the total transmittance (for $n\geq 6$), although the increase is very small.
Fourth, the gradient increases more significantly with increasing $n$ for the extensive distortion [Fig.~\ref{fig:Transpara}(e)] than for the compressive one [Fig.~\ref{fig:Transpara}(b)].
With these observations, we speculate that the gradient may be related to the bulk property of the MgO barrier.
Atomic positions in the MgO barrier in MTJs are usually different from those in bulk MgO because of the effect from Fe electrodes.
When we increase the barrier thickness in the MTJ, the electronic structure of the MgO layers approaches that of bulk MgO, which may result in an increase of the gradient.

\begin{figure}
    \centering
    \includegraphics[width=0.7\linewidth]{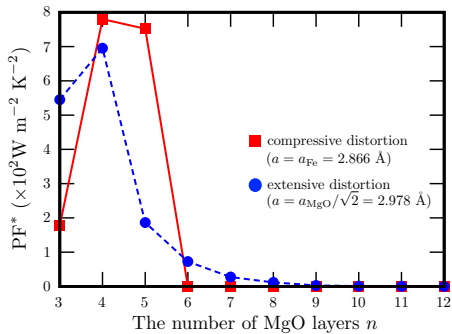}
    \caption{The power factor ($\text{PF}^*$) of the Fe(7ML)/MgO($n$ML)/Fe(7ML)(001) MTJ for the parallel magnetization configuration as a function of the MgO layers $n$ for the compressive ($a=a_\text{Fe}=2.866\text{ \AA}$, red solid curve) and the extensive ($a=a_\text{MgO}/\sqrt{2}=2.978\text{ \AA}$, blue dashed curve) distortions defined in Eq.~\eqref{eq:PF}.}
   \label{fig:PFpara}
\end{figure}

In addition to the Seebeck coefficient, we define the power factor to characterize the thermoelectric output power of the MTJ as
\begin{equation}
\text{PF}^* \equiv \frac{G}{A} S^2, \label{eq:PF}
\end{equation}
where $A$ is the cross section of the MTJ. 
In the definition, since there is usually no periodicity along the $z$ axis in MTJs, we use $G/A$, the conductance divided by the cross-sectional area of the MTJ, instead of the electrical conductivity, which is the analog of the resistance area product $RA$ and was used in Ref.~\onlinecite{Jia2016NJPFeMgO}.

The power factor of the Fe(7ML)/MgO($n$ML)/Fe(7ML)(001) MTJ (from $n=3$ to $n=12$) for the parallel magnetization configuration for the compressive and the extensive tetragonal distortions of the in-plane lattice parameter is shown in Fig.~\ref{fig:PFpara}.
For $n=4$ and $5$, the power factor for the compressive distortion ($a=a_\text{Fe}=2.866\text{ \AA}$) is larger than that for the extensive one ($a=a_\text{MgO}/\sqrt{2}=2.978\text{ \AA}$) because of the enhancement of the Seebeck coefficient for the compressive distortion (Fig.~\ref{fig:Seebeckpara}) caused by the interface resonant tunneling.
For $n\geq 6$, the power factor $\text{PF}^*$ suddenly decreases for both the compressive and the extensive distortions since the conductance decreases exponentially as the number of MgO layers $n$ increases.
Our present results for the Seebeck coefficient and the power factor indicate that the interface resonant tunneling can enhance the Seebeck coefficient and the power factor simultaneously for the compressive distortion with four or five atomic layers of the MgO barrier.

We here compare the power factor of the MTJ, $\text{PF}^*$, with the power factor of typical bulk thermoelectric materials. 
Since the definition of the power factor of bulk thermoelectric materials, $\text{PF}\equiv \sigma S^2$, includes the electric conductivity $\sigma$ [not the conductance $G$ in Eq.~\eqref{eq:G}], we need to set the total thickness of the MTJ along its stacking direction, denoted as $L_z$, to calculate the electric conductivity of the MTJ for the comparison. 
Using $L_z$, we calculated the electric conductivity of the MTJ as $\sigma = GL_z/A$, and utilized the relation $\text{PF} = L_z \text{PF}^*$ to obtain the value of $\text{PF}$ [see Eq.~\eqref{eq:PF}]. 
When $L_z = 100 \text{ nm}$, which is a typical length scale of real MTJs, the conventional power factor for our MTJ is about $\text{PF} \approx 8\times 10^{-5} \text{ W}\text{m}^{-1}\text{K}^{-2}$ using the maximum value of $\text{PF}^*$ (for $n = 4$ for the compressive tetragonal distortion), while the power factor of typical bulk thermoelectric materials is of the order of $10^{-5} - 10^{-3}  \text{ W}\text{m}^{-1}\text{K}^{-2}$\cite{Macia2015thermoelectric}.
%%%%%%%%%%%%%%%%%%%%%%%%%%%%%%%%%%%%%%%
%%%%%%%%%%%%%%%%%%%%%%%%%%%%%%%%%%%%%%%
%%%%%%%%%%%%%%%%%%%%%%%%%%%%%%%%%%%%%%%

\section{Discussion of the effect of nonlinearity, bias voltage and disorder}
So far, we have analyzed the Seebeck coefficient in the linear-response regime without considering bias voltage and disorder in order to understand its physical origin.
Here we discuss the effect of nonlinearity, bias voltage and disorder on the Seebeck coefficient.

Nonlinear Seebeck coefficients have been investigated in the research area of the transport in molecular junctions \cite{Zimbovskaya2017JChemPhysmolJunc} and strongly correlated quantum dots \cite{Azema2014PRBCondition}. 
Following the previous studies \cite{Azema2014PRBCondition, Zimbovskaya2017JChemPhysmolJunc}, we calculated a nonlinear Seebeck coefficient defined as $S = -\Delta \mu / (e\Delta T)$ with $\Delta \mu$ satisfying the open circuit condition, $I=0$, where we set $\mu_L =\mu_R+ \Delta \mu$ with $\mu_R = E_\text{F}$, as well as $T_\text{L} = T_\text{R} + \Delta T$ with $T_\text{R}  = 300 \text{ K}$ and $\Delta T = 30 \text{ K}$ [see Eq.~\eqref{eq:I}].
We found that the difference between the linear and nonlinear Seebeck coefficients is always less than $2 \ \mu\text{V}\text{K}^{-1}$.
We thus consider that the nonlinear effect in the Seebeck coefficients is negligible in our results.

Finite bias voltage has been known to break the symmetry between the left and right interface states, resulting in the suppression of the interface resonant tunneling \cite{Belashchenko2005PRBIRS, Rungger2009PRBBiasFeMgOFe}.
We therefore expect that it will suppress the enhancement of the Seebeck coefficient.
In order to confirm this expectation, we calculated the Seebeck coefficient using the transmittance under finite bias voltages taken from Ref.~\onlinecite{Masuda2017PRB}, in which the energy dependence of the transmittance of a Fe/MgO(5ML)/Fe(001) MTJ under various bias voltages was calculated. 
We found that the absolute value of the Seebeck coefficient decreases from $12.6$ to $4.1 \ \mu\text{V}\text{K}^{-1}$ with increasing the bias voltage from $0$ to $0.5 \text{ V}$, which can be attributed to the suppression of the interface resonant tunneling due to the applied bias voltage.

Disorder in MTJs will also affect the interface resonant tunneling and the Seebeck coefficient.
For example, in order to consider generic disorder theoretically, Rungger \textit{et al}.~\cite{Rungger2009PRBBiasFeMgOFe} added the small imaginary part to the energy when calculating the transmittance of a Fe/MgO(4ML)/Fe MTJ, which corresponds to the uniform level broadening. 
They found that the interface resonant tunneling in the minority-spin channel is suppressed with increasing the imaginary part, that is, increasing the effect of disorder. 
This result suggests that the enhancement of the Seebeck coefficient due to the interface resonant tunneling will be suppressed by the disorder.

\section{Summary and conclusion}
In the present work, we have calculated the Seebeck coefficient of the Fe(7ML)/MgO($n$ML)/Fe(7ML)(001) MTJ by means of the linear-response theory combined with a first-principles-based Landauer-B\"uttiker approach for several numbers of MgO layers $n$ (from $n=3$ to $n=12$) and for the compressive ($a=a_\text{Fe}=2.866$~\AA) and extensive ($a=a_\text{MgO}/\sqrt{2}=2.978\text{ \AA}$) tetragonal distortions of the in-plane lattice parameter.
We have found that the behavior of the Seebeck coefficient as a function of the number of MgO layers $n$ is completely different between the two kinds of distortions.
A detailed analysis of the transmittance has clarified that the interface resonant tunneling can enhance the Seebeck coefficient for the compressive lattice distortion with four or five atomic layers of the MgO barrier.
In addition to the Seebeck coefficient, we have calculated the power factor defined in Eq.~\eqref{eq:PF}.
It has been found that the power factor can additionally be enhanced for the compressive distortion due to the interface resonant tunneling.
These results indicate that control of the barrier thickness and the in-plane lattice distortion will be effective for maximizing the thermoelectric properties of MTJs.

\begin{acknowledgments}
The authors are grateful to Prof.~G.~E.~W.~Bauer, Dr.~R.~Iguchi, and Dr.~K.~Nawa for fruitful discussions and useful comments.
This work was supported by CREST “Creation of Innovative Core Technologies for Nano-enabled Thermal Management” (Grant No.~JPMJCR17I1) from JST, Japan and NIMS $\text{MI}^2\text{I}$. The crystal structures were visualized by using VESTA \cite{VESTA}. 
\end{acknowledgments}

\appendix*
\section{Antiparallel magnetization configuration}

\begin{figure}
    \centering
    \includegraphics[width=0.9\linewidth]{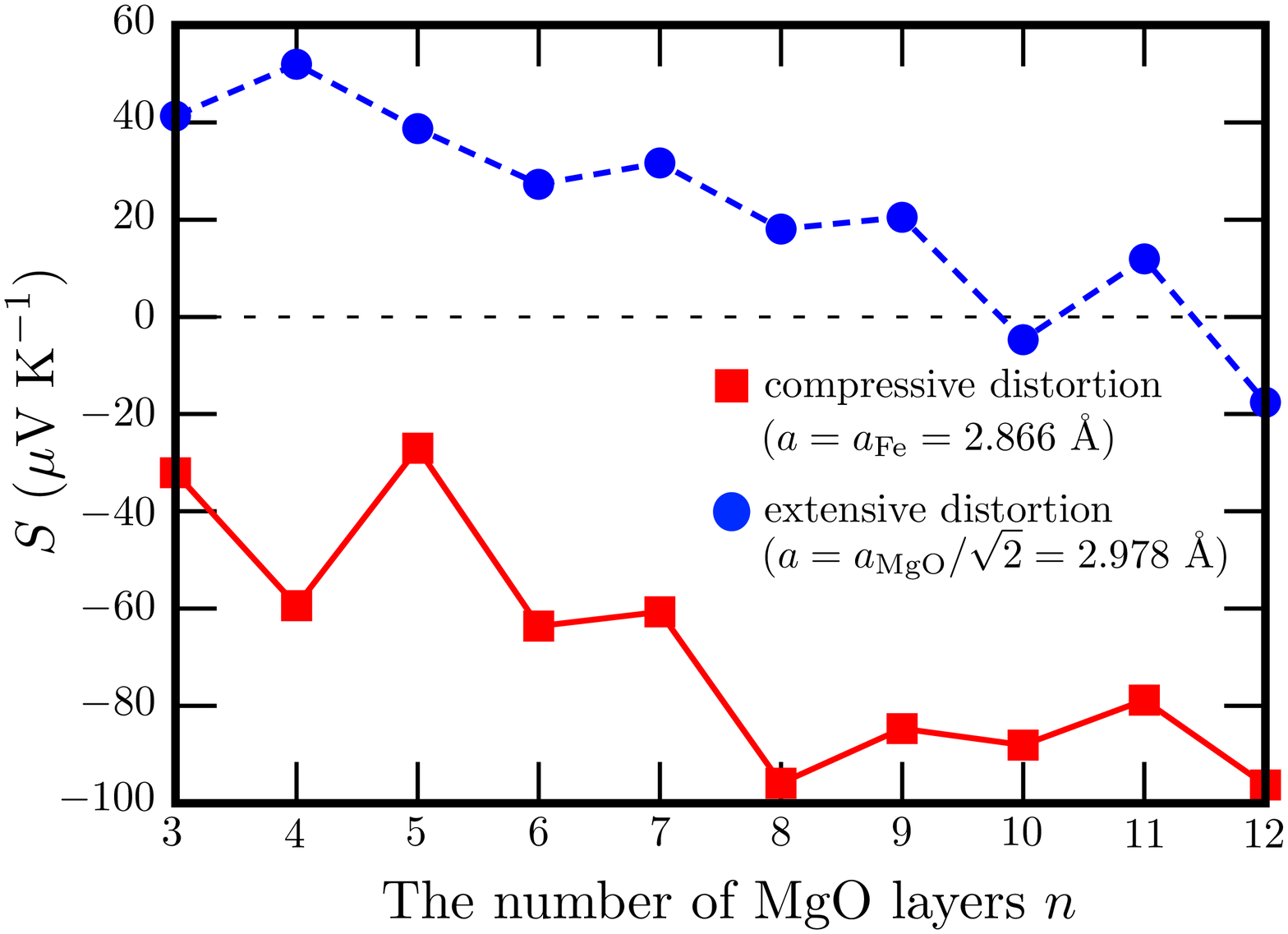}
    \caption{The Seebeck coefficient of the Fe(7ML)/MgO($n$ML)/Fe(7ML)(001) MTJ for the antiparallel magnetization configuration as a function of the number of MgO layers $n$ for the compressive ($a=a_\text{Fe}=2.866\text{ \AA}$, red solid curve) and the extensive ($a=a_\text{MgO}/\sqrt{2}=2.978\text{ \AA}$, blue dashed curve) distortions.}
    \label{fig:Seebeckanti}
\end{figure}

In this appendix, we analyze the Seebeck coefficient, the transmittance, and the power factor ($\text{PF}^*$) for the antiparallel magnetization configuration.
Figure~\ref{fig:Seebeckanti} shows the Seebeck coefficient of the Fe(7ML)/MgO($n$ML)/Fe(7ML)(001) MTJ for the antiparallel magnetization configuration as a function of the number of MgO layers $n$ for the compressive ($a=a_\text{Fe}=2.866 \text{ \AA}$) and the extensive ($a=a_\text{MgO}/\sqrt{2}=2.978 \text{ \AA}$) tetragonal distortions of the in-plane lattice parameter.
%The signs of them are different except for $n=10$ and $12$ unlike the case for the parallel magnetization configuration, which is the reason we do not show the absolute value but the value itself of the Seebeck coefficient.
With increasing the number of MgO layers $n$, the absolute value of the Seebeck coefficient for the compressive distortion increases, while the one for the extensive distortion decreases. 

\begin{figure}
    \centering
    \includegraphics[width=\linewidth]{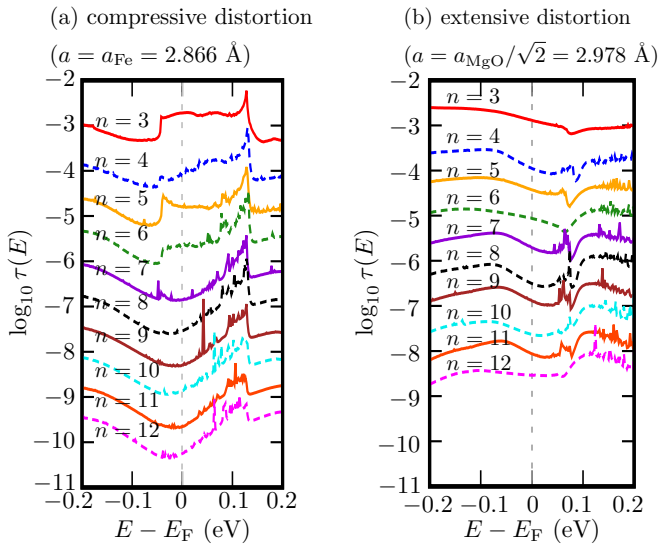}
    \caption{Total transmittances in the Fe(7ML)/MgO($n$ML)/Fe(7ML)(001) MTJ for (a) the compressive ($a=a_\text{Fe}=2.866\text{ \AA}$) and (b) the extensive ($a=a_\text{MgO}/\sqrt{2}=2.978\text{ \AA}$) distortions for the antiparallel magnetization configuration as a function of energy relative to the Fermi energy $E-\EF$ for the number of MgO layers $n$ from $n=3$ to $n=12$.}
    \label{fig:Transanti}
\end{figure}

In order to understand the behavior of the Seebeck coefficient, we plot the energy dependence of the total transmittances for both the distortions for the antiparallel magnetization configuration in Fig.~\ref{fig:Transanti}.
We show only the total transmittance in Fig.~\ref{fig:Transanti} because the transmittances in the majority- and minority-spin channels are almost equal in the antiparallel magnetization configuration.
For the compressive distortion [Fig.~\ref{fig:Transanti}(a)], we can see the enhancement around $E-E_\text{F} = [-0.05 \text{ eV}, 0.14 \text{ eV}]$ for $n=3$ that might originate from the interface resonant tunneling.
With increasing the number of MgO layers $n$, the transmittance around $E-E_\text{F} = [-0.05\text{ eV}, 0.05 \text{ eV}]$  decreases, while the transmittance around $E-E_\text{F} = 0.1 \text{ eV}$ hardly changes, resulting in increasing the asymmetric energy dependence of the transmittance with respect to the Fermi energy.
Therefore, the absolute value of the Seebeck coefficient increases with increasing $n$ for the compressive distortion (Fig.~\ref{fig:Seebeckanti}).

For the extensive distortion [Fig.~\ref{fig:Transanti}(b)], we can see the enhancement of the transmittance in $E-E_\text{F} \geq 0.1 \text{ eV}$ for all $n$ except for $n=3$.
While this produces the negative Seebeck coefficient, the complicated contributions from the band structure of Fe electrode provide the positive Seebeck coefficient through the energy dependence in $E-E_\text{F} \leq 0 \text{ eV}$.
These two contributions result in the dependence on $n$ of the Seebeck coefficient for the compressive distortion in Fig.~\ref{fig:Seebeckanti}.

\begin{figure}
    \centering
    \includegraphics[width=0.9\linewidth]{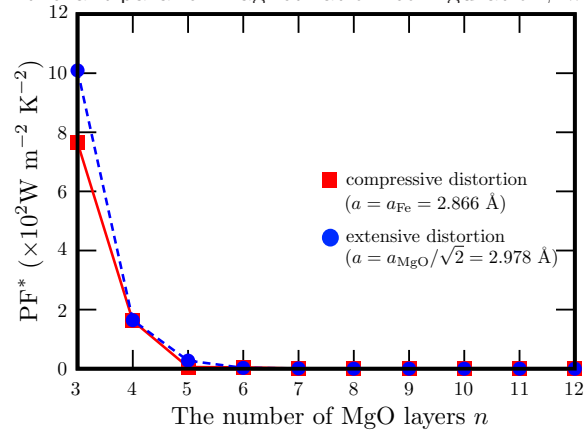}
    \caption{The power factor ($\text{PF}^*$) defined in Eq.~\eqref{eq:PF}, of the Fe(7ML)/MgO($n$ML)/Fe(7ML)(001) MTJ for the antiparallel magnetization configuration as a function of the MgO layers $n$ for the compressive ($a=a_\text{Fe}=2.866\text{ \AA}$, red solid curve) and the extensive ($a=a_\text{MgO}/\sqrt{2}=2.978\text{ \AA}$, blue dashed curve) distortions.}
  \label{fig:PFanti}
\end{figure}

We plot the power factor for the antiparallel magnetization configuration for the compressive and extensive distortions in Fig.~\ref{fig:PFanti}. 
Unlike the case for the parallel magnetization configuration (Fig.~\ref{fig:PFpara}), the power factor for the antiparallel magnetization configuration decreases monotonically and exponentially with increasing $n$.
This is due to the exponential decrease in the conductance for the antiparallel magnetization configuration, which is more rapid than that for the parallel magnetization configuration.

\providecommand{\noopsort}[1]{}\providecommand{\singleletter}[1]{#1}%

\end{document}